\documentclass[runningheads,10pt,a4paper]{llncs}
\usepackage{graphicx}
\usepackage{xcolor}
\usepackage{amsmath, amsfonts, amssymb,  enumerate}
\usepackage{tikz}
\usetikzlibrary{matrix}
\usetikzlibrary{positioning}
\usetikzlibrary{arrows}
\usepackage[linesnumbered, ruled]{algorithm2e}
\usepackage{mathrsfs}
\usepackage{multirow,array}

\usepackage{hyperref}

\newcommand{\g}{g}
\newcommand{\ignore}[1]{}
\newcommand{\eop}{{\hfill $\blacksquare$} }

\newcommand{\noi}{\noindent}

\newtheorem  {thm}{{\bf Theorem}}
\newtheorem  {cor}{{\bf Corollary}}
\newcommand{\TR}[0]{}

\begin{document}
\title{Financial replicator dynamics: emergence of systemic-risk-averting strategies} 
%
%
\author{Indrajit Saha  \and Veeraruna Kavitha} 
%
 
%
\institute{ IEOR, IIT Bombay, India \\
\email{\{indrojit, vkavitha\}@iitb.ac.in}}
\maketitle              
\begin{abstract}

We consider a random financial network  with a large number of agents. The agents connect through credit instruments borrowed from each other or through direct lending, and these create the liabilities. The settlement of the debts of various agents at the end of the contract period can be  expressed  as solutions of random fixed point equations. Our first step is to derive these solutions (asymptotically), using a recent result on random fixed point equations.
 We consider a large population in which the agents  adapt one of the two available  strategies, risky or risk-free investments,  with an aim  to maximize their   expected returns (or surplus).  We aim to study the emerging strategies when different types of replicator dynamics capture inter-agent interactions. We theoretically reduced the analysis of the complex system to that of an appropriate ordinary differential equation (ODE).  We proved that the equilibrium strategies converge almost surely to that of an  attractor of the ODE. We also derived the conditions under which a mixed evolutionary stable strategy (ESS) emerges; in these scenarios the replicator dynamics converges to an equilibrium at which the expected returns of  both the populations are equal.  
 Further the average dynamics (choices based  on large observation sample)  always averts systemic risk events (events with large fraction of defaults).  
  We verified through Monte Carlo simulations that the equilibrium suggested by the ODE method indeed represents the limit of the dynamics.
\keywords{Evolutionary  stable strategy (ESS), Replicator dynamics, Ordinary differential equation, Random graph,  Systemic risk, Financial network.}
\end{abstract}

\ignore{
Suppose that reasoning people do not choose strategies or behaviour in games, but instead are `hard-wired' by the players' genes. Suppose further that those strategies that are relatively successful (or rather, the genes associated with those strategies and behaviours) grow while less successful strategies die out. We might want to ask what strategies will be selected by such an evolutionary process. This question has led biologists to use game theory to study animal behaviour.
A related question concerns firms that compete in the market place. Perhaps the firms' policies are not choosen by sophisticated game theorists, but rather are associated with `rules of thumb'. In this case, those firms with rules of thumb that are worse (given the rules of thumb of the other firms) might go bankrupt,
leaving only a `population' of firms with more successful rules. Such competition might mimic rational choice in that the outcome might be that only `well-run' firms survive.}

\section{Introduction}

We consider a   financial network with   large number of agents. 
These agents are interconnected  to each other through financial  commitments  (e.g., borrowing,lending etc). In addition they make investments in either risk-free (risk neutral) or risky derivatives. 
In such a system the agents  not only face  random economic shocks (received via significantly smaller returns of their risky investments), they are also affected by the percolation of  the shocks faced by their neighbours (creditors), neighbours of their  neighbours etc. 
 In the recent years from $2007 -2008$ onwards,  there is a surge of activity to study the financial and systemic level risks  caused by such a percolation of shocks (\cite{allen2000financial,acemoglu2015systemic,eisenberg2001systemic,Systemicrisk}).
Systemic risk is the study of the risks related to  financial networks, when individual or entity level shocks can trigger   severe instability at system level that can collapse the entire economy (e.g.,   \cite{allen2000financial,acemoglu2015systemic,eisenberg2001systemic}). In this set of papers, the authors study the  kind of topology (or graph structure)  that is more stable towards the  percolation of shocks in  financial network, where stability is measured in terms of  the  total number of defaults in the network.  

In contrast to many existing studies in  literature related to systemic risk, we consider heterogeneous agents and we consider evolutionary framework.  
In our consideration, there are two  groups of agents   existing simultaneously  in the  network;  one group of agents invest in risk-free instruments,  while the   other group  
considers risky investments. The second group borrows   money from the other members of the network to gather more funds towards  the risky investments  (with much higher expected returns). 
These investments  are subjected to large (but rare) economic shocks,   
which can potentially percolate  throughout  the  network and can even affect the `risk-free' agents; the extent of percolation 
    depends upon   relative sizes of the two groups. We consider that new agents join such a network after each round of investment;  they choose their investment type (risky or risk-free) based on  their observations of the  returns  (the surplus of the agents after paying back their  liabilities) of a random sample of agents that invested  in   previous round.  The relative sizes of the two groups changes, the network structure changes,  which influences the (economic shock-influenced)  returns  of the agents in the next round, which in turn influences the decision of the new agents for the round after.   Thus the system evolves after each round. We study this evolution  process  using the well known evolutionary game theoretic tools.

In a financial network perspective, this type of work is new to the  best of our knowledge. We found few papers that consider evolutionary approach in  other aspects related to finance; in \cite{YangKe},  the authors study the financial safety net (a  series of the arrangement of the firms to maintain financial stability), and  analyze the   evolution of the bank strategies  (to take insurance or not); 
recently in  \cite{LiHonggang} authors consider an evolutionary game theoretic model with three types of players, i) momentum traders ii) contrarian traders iii) fundamentalists and studied the evolution of the relative populations. As already mentioned, these papers relate to very different aspects in comparison with our work. 

\noindent \underline{Evolutionary stable strategies:}
Traditionally evolutionary game models have been studied in the literature to study animal behaviour. The key ingredients of the evolutionary game models are a) a large number of players, b) the dynamics and c) the pay-off function (e.g., see the pioneering work \cite{Smith}).   
Replicator dynamics deals with evolution of strategies, reward based learning  in dynamic evolutionary games. Typically it is shown that these dynamics converge to a stable  equilibrium point  called Evolutionary Stable Strategy (ESS), which can be seen as a refinement of a strict Nash Equilibrium (\cite{Smith});  a  strategy prevailing in a large population  is called evolutionary stable if any small fraction of mutants  playing a different strategy get wiped out eventually.  Formally,   in a two player  symmetric game, a pure strategy $\hat{s}$ is  said to be evolutionary stable  if  
\begin{enumerate}
\label{Eqn_def_ESS}
 \item  $ (\hat{s},\hat{s})$ is a Nash equilibrium, i.e., 
  $u(\hat{s}, \hat{s}) \geq u( s^{'},\hat{s})$ $\forall s^{'}$ and, 
 \item   If $(\hat{s},\hat{s})$  is not a strict Nash equilibrium, i.e.,  $\exists$ some $s^{'} \neq \hat{s}$  such that $u( \hat{s}, \hat{s}) = u( s^{'}, \hat{s})$, then $u( \hat{s}, s^{'}) > u( s^{'}, s^{'})$.
\end{enumerate}

We study the possible emergence of   evolutionary stable strategies,  when people  choose  either a risky or a risk-free strategy;  the main difference being that the returns of either group are influenced by the percolation of shocks. The returns  of the portfolios depend further upon the percolation of shocks due to  layered structure of financial connections, and not just on the returns of the investments, i.e., not just on economic shocks. Our main conclusions are two fold;  a) when agents consider large sample of data for observation and learning, the replicator  dynamics can settle to a mixed ESS, at which the expected returns of the two the groups are balanced; b) in many other scenarios, through theoretical as well as simulation based study, we observed that the replicator dynamics converges to  one of the two strategies, i.e., to a pure ESS  (after completely wiping out the other group).

The analysis of these complex networks  (in each round) necessitated  the study of  random fixed point equations (defined sample path-wise in large dimensional spaces), which represent the clearing vectors of all the agents (\cite{Systemicrisk,allen2000financial,acemoglu2015systemic,eisenberg2001systemic} etc).  The study is made possible because of the recent result in \cite{Systemicrisk}, which  provided an asymptotically accurate  one dimensional  equivalent solution.

\section{Large Population Finance Network}
\label{sec_model}
We consider 
random graphs, where the edges represent the financial connection between the two nodes. Any two nodes are connected with probability $p_{ss} > 0$ independent of the others, but the weights on the edges depend on (the number of) neighbors.  This graph represents a large financial network where borrowing and lending are represented by the edges and the weights over them. 
The modeller may not have access to the exact  connections of the network,  but random graph model is a good approach to analyse such a complex system. In particular we consider the graphs that satisfy the assumptions of \cite{Systemicrisk}.  
  
  The agents are repeatedly investing in some financial projects. In each round of investment, the   agents borrow/lend from/to some   random subset of the agents of the network. Some of them may invest the remaining
  in a risk-free investment (which has a constant rate of interest $r_s$).  While the others invest the rest of their money in risky investments which have random returns; we consider a binomial model in which returns are high (rate $u$) with high probability $\delta$ and can have large shocks (rate $d$), but with small probability ($1-\delta$); it is clear that $d < r_s < u$.  We thus have two types of agents, we call the group that invests in risk-free  projects as `risk-free'  group  ($G_1$), the rest are being referred to as `risky' group ($G_2$).

   New agents join the network in each round of investment. 
  They  choose their investment type, either risk-free or risky,  for the first time based on the previous  experience of the network and continue the same choice for all future rounds of investment.  
  The new agents learn from network experience (returns of agents of the previous round of investments) and  choose  a suitable investment type, that can potentially give them good returns. The new agents either learn from the experience of a  random sample (returns of  two random agents) of the network or learn from a large number of agents. In the former case, their choice of investment type depends upon the returns of the random sample in the previous round. While in the latter case the decision can also depend on the average utility of each group of the agents, obtained after observing large number of samples.

{\bf Two strategies:} As mentioned before, there are two strategies available in the financial market. Risk-free agents  of $G_1$ use strategy 1;  these agents lend some amount of their initial wealth to other agents (of $G_2$) that are willing to borrow,  while the rest is invested in a government security, for example, bonds, government project etc.  Risky agents of $G_2$  are  adapting strategy $2$, wherein they borrow funds from the other agents and invest in risky security, for example, derivative markets, stocks, corporate loans etc.
These agents also lend to other agents of $G_2.$
Let  $\epsilon_t $ be the  fraction of the agents in  $G_1$ group and let $n(t)$ be the total number of agents in round $t$.   Thus the total number of agents (during round $t$) in group 1 equals  $n_1(t) := |G_1 |  =  n(t) \epsilon_t$ and  $n_2(t) := |G_2|  =  n(t) (1 -\epsilon_t )$.

We consider that one new agent is added in each round \footnote{this approach can easily be generalized to several other types of dynamics and we briefly discuss a few of them towards the end.}, 
and thus  size of the graph/network is increasing.  The agents are homogeneous, i.e., they reserve  the same wealth $w >0$ for investments (at the  initial investment  period) of each round. 
{\it Each round is composed of two time periods, the agents invest during the \underline{initial investment period} and they obtain their returns after some given time gap.} 
The two time period model is borrowed from  \cite{acemoglu2015systemic,eisenberg2001systemic,Systemicrisk} etc.
 The new agents make their choice for the next (and the future) round(s),  based on their observations of  these returns of the previous round.

{\bf Initial investment phases:}   During the initial investment phases (of any round $t$), any agent    $i\in G_{1}$  lends  to any agent $j \in G_2$ with probability $p_{ss}$ and it lends (same) amount\footnote{This normalization, (after choosing the required parameters, like $w$, appropriately) is done to derive simpler  final expressions. }   $w/(n(t) p_{ss})$ to each of the approachers based on   the number that approached it for loan;  let $I_{ij}$ be the indicator of this lending event.  
Note that for large $n(t)$, the number  of approachers of $G_2$ approximately equals $n(t) (1-\epsilon_t) p_{ss}$, and, 
thus any agent of $G_1$ lends approximately  $w (1-\epsilon_t)$ fraction  to agents of $G_2$.   The agents of $G_1$  invest the rest $w\epsilon_t$ in risk-free investment (returns with fixed rate of interest $r_s$).

Let   $\tilde{w}$ be the accumulated wealth\footnote{These amounts could be random and different from agent to agent, but with large networks (by law of large numbers) one can approximate  these to be constants.} of any agent of  $G_2$ out of which a positive fraction $\alpha $  is invested towards the other banks of $G_2$ and $(1-\alpha)$ portion is invested in risky security.
\noindent Thus   the accumulated wealth of a typical $G_2$ agent is  governed by the following equation: 
\begin{equation}
\tilde{w}  = \underbrace{w+ w\epsilon }_\text{Initial wealth + Borrowed from $G_1$}+\underbrace{\tilde{w} \alpha}_\text{Lend/borrow $G_2$}  \mbox{ and thus }  \tilde{w} = \frac{w(1+\epsilon)}{ (1-\alpha)}.
\end{equation}
 Thus the total  investment towards the risky venture equals  $\tilde{w} (1-\alpha)= w(1+\epsilon)$.
The $G_2$ agents have to settle their liabilities at the end of the return/contract period (in each round) and  this would depend upon their returns from the risky investments.   Thus the total liability of any agent of  $G_2$ is  $y = (w\epsilon +\tilde{w}\alpha )(1+r_b)$, where $r_b$ is  the borrowing rate\footnote{For simplicity of explanation, we are considering constant terms to represent all these quantities, in reality they would be i.i.d. quantities which are further independent of other rounds and the asymptotic analysis would go through as in \cite{Systemicrisk}.}; by  simplifying 
$$ y= \frac{w(\epsilon +\alpha)(1+r_b)}{( 1-\alpha)}. $$
Similarly, any agent of $G_2$ lends the following amount to each of its approachers (of $G_2$):
\begin{equation}
\label{Eqn_liabOne}
\frac{\alpha {\tilde w} }{n(t) (1-\epsilon_t) p_{ss}} = \frac{ \alpha w(1+\epsilon)}{n(t) (1-\epsilon_t) p_{ss}  (1-\alpha)} .
\end{equation}

\begin{figure}
\vspace{-1.9mm}
\hspace{-12mm}
\begin{minipage}{7cm}
    \centering
     \includegraphics[width =0.45\textwidth]{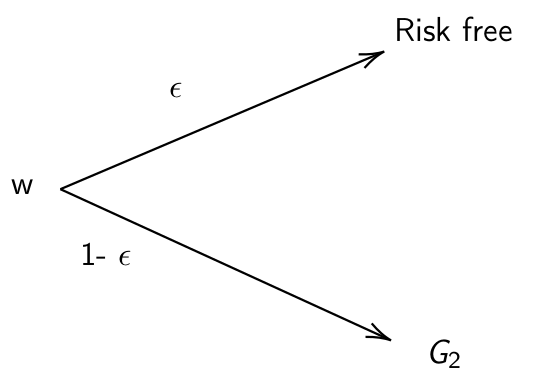}
       \caption{Apportioning of $G_1$}
    \label{figure1}
\end{minipage}
 \begin{minipage}{7cm}
    \centering
     \includegraphics[width =0.55\textwidth]{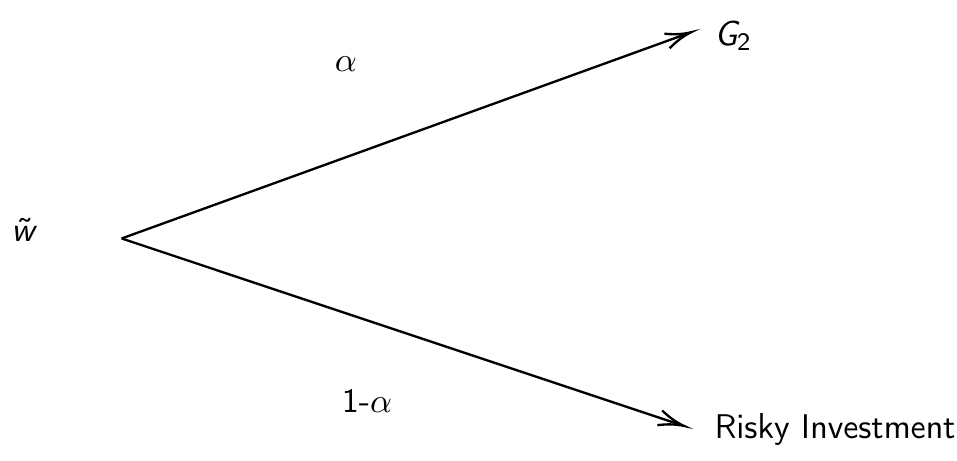}
     \caption{Apportioning of  $G_2$}
    \label{figure2}
    \end{minipage}
    \vspace{-6mm}
\end{figure}

{\bf Return and Settling phases,  Clearing Vectors:}
\ignore{ The banks have heterogeneous believe that the future value of the asset price will go up.  Therefore some of the banks are borrowing more and invest the money for a higher return while the other group of banks make a less risky investment. Although the channel of risk for the $G_{1}$ banks is inherent because the failure of the group $G_{2}$ banks can cause the portfolio value of the banks goes down.
}
We fix the round $t$ and avoid notation $t$ for simpler notations. The agents of $G_2$ have to clear their liabilities during this phase in every round.  
Recall the agents of $G_2$ invested $w(1+\epsilon)$ amount in risky-investments and the corresponding random returns (after economic shocks) are:
\begin{equation}
    K_i=
    \begin{cases}
      w(1+\epsilon)(1+u)=:k_u, & \text{w.p. (with probability) }\ \delta \\
      w(1+\epsilon)(1+d)=: k_d, & \text{otherwise}
    \end{cases}
  \end{equation}
  This is the well known binomial model, in which the upward moment occurs with probability $\delta$ and downward moment with $(1-\delta).$
 The agents have to return $y$ (after the interest rate $r_b$) amount to their creditors, however may not be able to manage the same because of the above economic shocks.  
In case of default,  the agents return the maximum possible;
let $X_i$ be the amount cleared by   the $i$-{th agent of  group $G_2$.  
 Here we consider a standard   bankruptcy rule, limited liability and pro-rata basis repayment of the debt contract (see  \cite{acemoglu2015systemic,eisenberg2001systemic}), where the amounts returned are proportional to their liability ratios.
 Thus  node   $j$ of $G_2$  pays back $X_j L_{ji}/ y$  towards node $i$,  where  $L_{ji}$   the  amount borrowed  (liability) during initial investment phases equals (see the details of previous subsection and equation (\ref{Eqn_liabOne})):
 
\begin{equation}
    L_{ji}=
    \begin{cases}
      I_{ji} \frac{w}{n p_{ss}}, & \text{if}\ i \in G_1 \\
      I_{ji} \frac{\alpha w(1+\epsilon)}{n p_{ss} (1-\alpha)(1-\epsilon)}, & \text{if}\ i \in G_2.
    \end{cases} 
  \end{equation}
Thus the maximum amount cleared by any agent $j \in {\cal G}_2$, $X_j$,  is given by the following  fixed point equation in terms of the clearing vector 
$\{ X_i\}_{i \in G_2}$ composed    of clearing values of all the agents
  (see \cite{acemoglu2015systemic,eisenberg2001systemic} etc):
 
\begin{equation}
\label{Clearing vector}
 X_i =\min \left  \{  \bigg( K_i+ \sum_{j\in G_2} X_j\frac{L_{ji}} {y} - v \bigg )^+, \    y \right \},
\end{equation}

with the following details: the term $K_i$ is  the return of the risky investment,  the term     $ \sum_{j\in G_2} X_j\  L_{ji}/y $  equals the claims form the other agents (those borrowed from agent $i$) and $v$ denotes the  taxes to pay.  In other words, agent $i$  will
pay back  the (maximum possible) amount  $ K_i+   \sum_{j\in G_2} X_j\frac{L_{ji}}{y} - v$  in case of a default,  and  in the other event, will exactly pay back the  liability amount $y$.
 
\underline {Surplus} of any agent  is defined as  the amount obtained from various investments, after clearing all the liabilities.  This represents the utility of the agent in the given round. 
The surplus  of the agent  $i \in G_{2}$: 
\begin{equation}
\label{Retun for G_2}
R^2_i = \left (K_i+\sum_{j\in G_2} X_j\frac{L_{ji}}{y} -v-y \right )^+,
\end{equation}
while that of agent  $i \in G_1$ is  given by:
\begin{equation}
\label{Retun for G_1}
R^1_i =  \left  ( w\epsilon(1+r_s)+\sum_{j\in G_2} X_j\frac{L_{ji}}{y}  -v \right )^+.
\end{equation}
In the above, the first term is the return from the    risk free investment. The second term  equals the returns  or claims form  $G_2$ agents (whom they lent) and $v$ denotes the amount of taxes.
 
\section{Asymptotic approximation of the  large networks}
We thus have dynamic graphs whose size  increases   with  each round.  In this section, we obtain appropriate asymptotic analysis of   these graphs/systems, with an aim to derive the pay-off of each group after each round. Towards this, we derive the (approximate) closed form expression of the equations (\ref{Retun for G_2}) and (\ref{Retun for G_1}), which are nothing but the per-agent returns after the settlement of the liabilities.

The returns of the agents depend upon  how other agents settle their liabilities to their connections/creditors.  Thus our first step is to derive the  solution of the clearing vector fixed point equations (\ref{Clearing vector}). Observe that the  clearing vector  $\{X_j\}_{j \in G_2}$ is the solution of the  vector-valued  random fixed point equations (\ref{Clearing vector})  in $n$-dimensional space (where $n$ is the size of the network), defined sample-path wise.

\underline{Clearing vectors using results of \cite{Systemicrisk}:} Our  financial framework can be analysed using the results of   \cite{Systemicrisk},  
as the details of the model match\footnote{Observe that $\alpha(1+\epsilon)/(\alpha+\epsilon) < 1$.} the assumptions of the paper. By \cite[Theorem 1]{Systemicrisk}, 
 the aggregate claims   converge almost surely  to constant values (as the network size increases to infinity): 
 
\begin{eqnarray*}
 \mbox{(claims of agents of  $G_1$)},  &  \displaystyle\sum_{j \in G_2}X_j \frac{L_{ji }}{y}  \to  & \frac{(1-\alpha)(1-\epsilon)}{\alpha+\epsilon} {\bar x}^{ \infty}  \mbox{ a.s.}, \mbox{ and }  \\
  \mbox{(claims of agents of $G_2$)},  & \displaystyle \sum_{j \in G_2}X_j \frac{L_{ji }}{y}  \to  & \frac{\alpha (1+\epsilon)}{(\alpha +\epsilon)} {\bar x}^{ \infty} \mbox{ a.s.},
\end{eqnarray*}
where  the common expected clearing value ${\bar x}^{ \infty}$ satisfies the following fixed point equation in one-dimension:
\begin{equation}
\label{Eqn_aggregate}
{\bar x}^{ \infty} = E \bigg[\min \left  \{  \bigg( K_i+  \frac{\alpha (1+\epsilon) }{\alpha + \epsilon}  {\bar x}^{ \infty}  - v\bigg)^+,  y  \right \}\bigg].
\end{equation}
Further by the same Theorem, the clearing vectors converge  almost surely to (asymptotically independent) random vectors:
\begin{equation}
 X_i \to  \min \left  \{  \bigg( K_i+  \frac{\alpha (1+\epsilon) }{\alpha + \epsilon}  {\bar x}^{ \infty}  - v \bigg)^+,  y \right \}, \mbox{ for  } i \in G_2. \end{equation}

By virtue of the above results,  the random returns given by equations (\ref{Retun for G_2}) and (\ref{Retun for G_1}),  converge  almost surely:

\begin{eqnarray}
\label{Return_1}
 R^1_i &\to&  \left  ( w\epsilon(1+r_s)+ \frac{(1-\alpha)(1-\epsilon)}{(\alpha +\epsilon)}  {\bar x}^{ \infty}  -v \right )^+, \mbox{ for each }  i \in G_1
\\
\label{Return_2}
 R^2_i & \to & \left  (K_i+\frac{\alpha (1+\epsilon) }{\alpha + \epsilon}  {\bar x}^{ \infty} -v-y \right )^+,  \mbox{ for each }  i \in G_2.
 \end{eqnarray}
\underline{Probability of Default} is defined as  the fraction of agents of $G_2$ that failed to pay back their full liability,  i.e.,  $P_d:= P({ X_i} <y)$. 
For large networks (when the initial network size $n_0 $ itself is sufficiently large), one can use the above approximate expressions and using the same we obtain the  default probabilities and the aggregate clearing vectors in the following (proof in Appendix). 
\begin{lemma}
\label{Lem_Average_clearing}The asymptotic  average
clearing vector and the  default probability of $G_2$ with $k_d >v$ is  given by:

\begin{equation}
\label{Eqn_eps_rnd_star}
(  {\bar x}^{ \infty}, P_d)=
    \begin{cases}
    (y,                                                                                                     \hspace{24mm} 0)                & \text{if }\ \  c_\epsilon  > \frac{y-\underline{w}}{y} \\
  \left   (\frac{\delta y +(1-\delta) \underline{w}}{1-(1-\delta)c_\epsilon},     \hspace{4mm}  1-\delta \right )     & \text{if} \ \  \frac{y-\overline{w}}{y-(1-\delta)(\overline{w}-\underline{w})} < c_\epsilon <\frac{y-\underline{w}}{y} \\
      \left (\frac{E[W]}{1-c_\epsilon},           \ \ \ \ \       \   1 \right )                                  & \text{if }  \ \ c_\epsilon < \frac{y- \overline{w}}{y- (1-\delta)(\overline{w}-\underline{w})}
    \end{cases}
  \end{equation}
  where, $c_\epsilon = \frac{\alpha +\alpha \epsilon}{\alpha +\epsilon}$,
$E[W] = \delta k_u + (1-\delta) k_d -v$ , $\underline{w} = k_d -v$ and $\overline{w}= k_u -v$. 
\eop
\end{lemma}
 \underline{Expected Surplus:} 
 By   virtue  of the  Theorem  developed in \cite[Theorem 1]{Systemicrisk}  we have a significantly simplified limit system, whose performance is  derived in the above lemma. We observe that this approximation is sufficiently close (numerical simulations illustrate good approximations), and assume the following as the pay-offs  of each group after each round of the  investments:
 \begin{align}
  & \phi_1 (\epsilon) : = E[R^1_i] =  \bigg(w\epsilon (1+ r_s) + \frac{(1-\alpha)(1-\epsilon)}{\alpha +\epsilon}{\bar x}^{ \infty} - v \bigg)^+,  \mbox{ for any agent of } G_1 \nonumber \\
  & \phi_2(\epsilon) : = E[R^2_i] =  E \bigg (K_i + \frac{(1+\epsilon)\alpha}{\alpha +\epsilon}{\bar x}^{ \infty} - v -y\bigg)^+ ,  \label{Eqn_phis}\\
  & = \bigg ( k_u + \frac{\alpha ( 1+\epsilon)}{\alpha +\epsilon } {\bar x}^{ \infty} - v-y\bigg)^+ \delta + \bigg ( k_d + \frac{\alpha ( 1+\epsilon)}{\alpha +\epsilon } {\bar x}^{ \infty} - v -y\bigg )^+(1- \delta), \nonumber
\end{align}for any agent of $G_2.$
 Observe here that the aggregate limits are almost sure constants, hence the expected surplus of all  the agents  of   the same group are equal, while the random returns of the same group are i.i.d. (independent and identically distributed). 
  
 \ignore{

Avg model

.. Table ...

Random Model 

Table ..
}

\section{Analysis of Replicator dynamics}

In every round of investments, we have a new network that represents the liability structure of all the agents of that round formed by the investment choices of  the agents,  and,  in the previous two sections we computed the (asymptotically approximate) expected returns/utilities of each agent of the network. 
As already mentioned in Section \ref{sec_model},  new agents join  the network in each round,  and choose their strategies  depending upon their observations of these expected returns of   the previous round.   

These kind of dynamics is well described in literature by name replicator dynamics (e.g., \cite{MeanWireless,LiHonggang,ESS} etc).
The main purpose of such a study is to derive  asymptotic analysis and answer some or all of the following questions:   will the dynamics converge, i.e., would the relative fractions of various populations  settle as the number of rounds increase? 
will some of the strategies disappear  eventually? if more than one population type survives what would be the asymptotic fractions? etc.  
 These kind of analysis are common in other types of networks (e.g., wireless networks (e.g., \cite{MeanWireless}), biological networks (e.g., \cite{ESS})), but are relatively less studied in the context of financial networks (e.g., \cite{LiHonggang}).  We are interested in knowing the asymptotic outcome of these kind of dynamics (if there exists one) and study the influence of various network parameters on the outcome. 
We begin with precise description of the two types of dynamics considered in this paper.

\subsection{Average Dynamics}
The new agent  contacts two random (uniformly sampled) agents    of the previous round.   If both the contacted agents belong to the same group, the new agent adapts the strategy of that group. When it contacts agents from both the groups it investigates more before making a choice; the new  agent observes significant  portion of the network, in that,  it  obtains a good estimate of  the average utility of agents belonging to both the groups. It adapts the strategy of the group with maximum (estimated) average utility.

Say it observes the average of each group with an error that is normally distributed with mean equal to  the expected  return of  the group and variance proportional to the size of the group, i.e., 
it observes (here ${\cal N}(0, \sigma^2)$ is a zero mean Gaussian random variable  with variance $\sigma^2$)
$$
{\hat \phi}_i (\epsilon) =\phi_i (\epsilon) + {\cal N}_i  \mbox{ with } {\cal N}_1  \sim  {\cal N} \left ( 0,  \frac{1}{{\bar c} \epsilon } \right  ) \mbox{ and }  {\cal N}_2  \sim  {\cal N} \left ( 0,  \frac{1}{  {\bar c} (1- \epsilon )  } \right ), 
$$
for some ${\bar c}$ large. 
Observe by this modeling that:  {\it the expected values of the observations are given by   
$(\phi_1(\epsilon), \ \phi_2 (\epsilon) )$ and are determined by the relative proportions of the two populations, while the variance of any group reduces as its proportion increases to 1 and increases as the proportion reduces to zero.} We also assume that the estimation errors $\{ {\cal N}_1, {\cal N}_2 \}$ (conditioned on the relative fraction, $\epsilon$) corresponding to the two groups are independent. 
Then the probability that the new agent chooses strategy 1 is given by
$$
Prob (  {\hat \phi}_1 (\epsilon) - {\hat \phi}_2 (\epsilon)   > 0  ) =  Prob (  {\cal N}_2  - {\cal N}_1 \le    \phi_1 (\epsilon) - \phi_2 (\epsilon) ) ,
$$
which by (conditional) independence of Gaussian random variables equals\footnote{  because $\frac{1}{\epsilon} + \frac{1} {1-\epsilon} =  \frac{1}{\epsilon (1-\epsilon)} $ }
\begin{eqnarray}
\label{Eqn_g_epsilon}
g(\epsilon) :=  \int_{-\infty}^{  \left ( \phi_1 (\epsilon ) -  \phi_2 (\epsilon) \right )  \sqrt{{\bar c} \epsilon (1-\epsilon) }   }  e^{- x^2 / 2}  \frac{dx }{\sqrt{2 \pi}} . 
\end{eqnarray}

Let   $( n_1(t), n_2(t))$  respectively represent the sizes of  $G_1$ and $G_2$ population after round $t$ and note that   $\epsilon_t  = \frac{n_1(t)}{n_1(t)+ n_2(t)} $.  Then the   system dynamics is given by the following ($g(\cdot)$ given by (\ref{Eqn_g_epsilon})):
 
 \vspace{-4mm}
 {\small \begin{eqnarray} 
    (n_1(t+1)  , \    n_2(t+1))  &=& \left \{ 
    \begin{array}{llll}
    \big ( n_1(t)  + 1 ,  & n_2(t)    \big ) & \text{w.p.}\ \ \epsilon_t^2  + 2 \epsilon_t (1-\epsilon_t)  g (\epsilon_t) \\
  \big (n_1(t) ,  & \hspace{-2mm} n_2(t)  + 1   \big )  & \text{w.p.} \ \  (1- \epsilon_t )^2 + 2 \epsilon_t (1-\epsilon_t) (1- g (\epsilon_t) ) . \\
    \end{array}  \right .  \nonumber  \\
      \label{Eqn_repl_avg_dynamics}
  \end{eqnarray}}
   It is clear that (with $\epsilon_0$ and $n_0$ representing the initial quantities),   
\begin{eqnarray*}
\epsilon_{t+1} &=& \frac{n_1(t+1)}{t+n_0+1} =  \frac{(t+n_0) \epsilon_t + Y_{t+1} }{t+n_0+1} = \epsilon_t + \frac{1}{t+ n_0+1} \left ( Y_{t+1} -\epsilon_{t} \right )  \mbox{ where, } \\  
    Y_{t+1}  &=&
    \begin{cases}
    1 & \text{wp}\ \ \epsilon_t^2 + 2 \epsilon_t (1-\epsilon_t)  g (\epsilon_t)  \\
   0 & \text{wp} \ \ (1- \epsilon_t )^2 + 2 \epsilon_t (1-\epsilon_t) (1- g (\epsilon_t) ) \mbox{, for all }  t\ge 1.
    \end{cases}
  \end{eqnarray*}
  One can rewrite the update equations as 
\vspace{-2mm}
  
{\small 
 \begin{eqnarray*}
\epsilon_{t+1} &=&   \epsilon_t + \frac{1}{t+n_0+1} \left ( h (\epsilon_t) + M_{t+1}  \right ),  \mbox{ with,  } \ \
M_{t+1} \ := \  Y_{t+1} -\epsilon_{t} - h(\epsilon_t),  \mbox{ where, }  \\
 h(\epsilon) & := &   E \big [ Y_{t+1}- \epsilon_t  | \epsilon_t = \epsilon  \big ]  = \epsilon (1-\epsilon) \left ( 2 g (\epsilon)  - 1 \right ) \mbox{ for any }  0 \le \epsilon \le 1.
    \end{eqnarray*}}
  and observe that  (with ${\cal F}_t$ the natural filtration of the process till $t$),
  \begin{eqnarray*}
  E[ M_{t+1}  | {\cal F}_t ]  &=&  E[ M_{t+1}  |  \epsilon_t ]  = 0  \mbox{ and }\ 
  E[ M_{t+1}^2  | {\cal F}_t ]  \ \le \  C \mbox{ for some constant }  C < \infty. 
    \end{eqnarray*}
Further observe that 
$$ \hspace{7mm}
0 \le  \epsilon_t \le 1  \mbox{ for all }  t \mbox{ and all sample paths}. 
$$  
Thus our algorithm satisfies assumptions\footnote{The assumptions require that the process is defined for the entire real  line. One can easily achieve this by letting $h(\epsilon) = 0$ for all $\epsilon \notin [0,1]$, which still ensures required Lipschitz continuity and by extending $M_{t+1} = 0$ for all $\epsilon_t \notin [0,1]$.} {\bf A}.1 to {\bf A}.4 of \cite{Borkar}  and hence we have using \cite[Theorem 2]{Borkar} that 
\begin{thm}
\label{Thm_avg_conv}
The sequence $\{\epsilon_t\}$ generated by average dynamics  (\ref{Eqn_repl_avg_dynamics}) converges almost surely (a.s.)  
to a (possibly sample path dependent) compact connected internally chain transitive invariant set of ODE:  
\begin{equation}
\dot{\epsilon_t}  = h(\epsilon_t) .   \hspace{ 20mm}  \mbox{ \eop}
\label{Eqn_Avg_ODE}
\end{equation}
\end{thm}

The dynamics  start with   initial condition $\epsilon_0 \in (0, 1)$
and clearly would remain  inside the interval $[0,1]$, i.e., $\epsilon_t \in [0,1]$ for all $t$ (and almost surely).  Thus we consider the invariant sets of ODE \eqref{Eqn_Avg_ODE} within this interval   for some interesting case studies in the following (Proof in Appendix).
 \begin{cor}
 \label{Lem_avg_dyn}
 Define ${\bar r}_r  := u \delta + d (1-\delta)$. And assume $w (1+d) > v$ and observe that  $u > r_b  \ge r_s > d.$
 Assume $\epsilon_0 \in (0, 1).$
 Given the rest of the parameters of the problem,  there exists a ${\bar \delta} < 1$   (depends upon the instance of the problem) such that the following statements are valid for all $\delta \ge {\bar \delta}$: 
 \begin{enumerate}[(a)]
 \item If  ${\bar r}_r  >  r_b > r_s$ then  $\phi_2 (\epsilon) > \phi_1 (\epsilon) $ for all $\epsilon$, and  $\epsilon_t \to  0 $ almost surely.  
 \item  If  $\phi_1 (\epsilon) > \phi_2 (\epsilon) $ for all $\epsilon$   then $\epsilon_t \to 1 $ almost surely. 
   \item When $ r_b > {\bar r}_r  >  r_s$,   and case (b) is negated    there exists a unique zero $\epsilon^*$ of the equation $\phi_1(\epsilon) - \phi_2(\epsilon) = 0$  and  
   $$ \hspace{10mm} \epsilon_t \to  \epsilon^{*}  \mbox{  almost surely;   further  for $\delta \approx 1$,  }  \epsilon^{*}  \approx   \frac {  r_b - {\bar r}_r  } {  {\bar r}_r  - r_s } .  \hspace{15mm}   \mbox{\eop}
   $$   
 \end{enumerate}
 \end{cor}
 From (\ref{Eqn_phis}) and Lemma \ref{Lem_Average_clearing},
 it is easy to verify that all the limit points are evolutionary stable strategies (ESS). 
 Thus the replicator dynamics either settles to a pure strategy ESS or mixed ESS (in part (c) of the corollary),  depending upon the parameters of the network;
  after a large number of rounds, either the fraction of agents following one of the strategies converges to one or zero  or  the system reaches a mixed ESS which balances the expected returns of the two groups.

In many scenarios, the expected rate of return of the risky investments is much higher than  the rate of interest related to lending/borrowing, i.e., ${\bar r}_r > r_b$. Further the assumptions of the corollary are satisfied by more or less all the scenarios (due to standard no-arbitrage assumptions) and because the shocks are usually rare (i.e., $\delta $ is close to 1).  Hence by the above corollary, in 
majority of scenarios, the average dynamics converges to a pure strategy with all `risky' agents (i.e., $\epsilon_t\to 0$). The group $G_1$ gets wiped out and almost all agents invest  in risky ventures, as the expected rate of returns is more even in spite of large economic shocks.  One can observe a converse or {\it a mixed ESS    when the magnitude of the  shocks  is   large ($d$ too small)  or when the shocks are too often to make ${\bar r}_r < r_b$. }

 \subsection{Random dynamics}
 When the new agent contacts two random agents  of different groups,   its choice depends directly upon the returns of the two contacted agents. The rest of the details remain the same as in average dynamics.  In other words,  the new agents observe less,  their investment choice is solely based on the (previous round)  returns  of the two contacted agents.   In this case the dynamics are governed by the following (see \eqref{Return_1}-\eqref{Return_2}):

\begin{eqnarray}
\label{Eqn_Repl_rand_dyan}
    (n_1(t+1) ,n_2(t+1))  &=&
    \begin{cases}
    \left ( n_1(t)  + 1 ,  \hspace{10mm} n_2(t) \right ) & \text{wp}\ \ \epsilon_t^2 \\
    \left (n_1(t) ,   \hspace{16mm} n_2(t)  + 1 \right )  & \text{wp} \ \ (1- \epsilon_t )^2 \\
     \left (n_1(t) + G (\epsilon_t )  ,  \ \  n_2(t) + (1-  G (\epsilon_t)  ) \right )   & \text{else, with}  \\
    \end{cases} \nonumber  \\ 
    G (\epsilon_t)  &=&  1_{\lbrace R_i^1 \ge  R_i^2  \rbrace} \\
    & =&   1_{  \left \{  \left (  w\epsilon(1+r_s)+ \frac{(1-\alpha)(1-\epsilon)}{(\alpha +\epsilon)}  {\bar x}^{ \infty}  -v  \right )^+ \ge   \left (K_i+\frac{\alpha (1+\epsilon) }{\alpha + \epsilon}  {\bar x}^{ \infty} -v-y \right )^+  \right \}  }, \nonumber 
  \end{eqnarray} where  $ {\bar x}^{ \infty} =  {\bar x}^{ \infty} (\epsilon_t)$ is  given by Lemma \ref{Lem_Average_clearing}.  Here we assume people prefer risk-free strategy under equality, one can easily  consider the other variants.
  Once again this can be rewritten as 
  \begin{align}
  \label{Eqn_repl_rand_dynamics}
   &\epsilon_{t+1}  =   \epsilon _{t} +\frac{Z_{t+1}-\epsilon_t}{t+n_0+1} \mbox{ with }  \ \ 
    Z_{t+1} =
    \begin{cases}
    1 & \text{wp}\ \ \epsilon_t^2 \\
   0 & \text{wp} \ \ (1- \epsilon_t )^2 \\
     G(\epsilon_t )   & \text{else}  .
    \end{cases}
  \end{align}
  
As in previous section the above algorithm satisfies   assumptions\footnote{In the current paper, we consider scenarios in which $h_R(\cdot)$ is Lipschitz continuous, basically under the conditions of    Corollary \ref{Cor_rand}. } {\bf A}.1 to {\bf A}.4 of \cite{Borkar}  and  once again using   \cite[Theorem 2]{Borkar}, we have:
\begin{thm}
\label{Thm_rand_conv}
The sequence $\{\epsilon_t\}$ generated by average dynamics  (\ref{Eqn_Repl_rand_dyan}) converges almost surely (a.s.)  
to a (possibly sample path dependent) compact connected internally chain transitive invariant set of ODE:  
\begin{equation}
\dot{\epsilon} (t)  = h_R (\epsilon (t) ),  \  h_R (\epsilon) :=  E_{\epsilon}\big [  Z_{t+1}- \epsilon_t | \epsilon_t = \epsilon \big ] =  \epsilon  (1-\epsilon) ( 2  E[G (\epsilon) ] - 1). \label{Eqn_Rand_ODE}
    \mbox{ \eop}
\end{equation}
\end{thm}
One can derive the analysis of this dynamics in a similar way as in average dynamics, however there is an important difference between the two dynamics; we can never have random dynamics converges to an intermediate  attractor, like the attractor in part (c) of Corollary \ref{Lem_avg_dyn} (unique 
$\epsilon^*$ satisfying $\phi_1 = \phi_2$). This is because  $E_\epsilon[G] = P( R^1 (\epsilon) > R^2 (\epsilon) )$ equals $0$, $1-\delta$ or 1 and never $1/2$  (unless $\delta = 1/2$, which is not a realistic case). Nevertheless,  we consider the invariant sets (corresponding to pure ESS)  within $[0,1]$ for some cases (Proof in Appendix): 
 \begin{cor}
 \label{Lem_random_dyn}
\label{Cor_rand}
 Assume $\epsilon_0 \in (0, 1).$  Given the rest of the parameters of the problem,  there exists a $1/2<  \delta < 1$   (depends upon the instance of the problem) such that the following statements are valid:
 
 \begin{enumerate}[(a)]
 \item If  $E_\epsilon[G]  =  0$   for all $\epsilon$
  or  $ 1-\delta$  for all $\epsilon$, then   $\epsilon_t \to  0 $ almost surely.  
 \item   If  $E_\epsilon[G]  =  1$  for all $\epsilon$, then   $\epsilon_t \to  1 $ almost surely.
  \item When $w(1+d) > v$ and $u > r_b \ge r_s > d$,  there exists a ${\bar \delta} < 1$ such that for all $\delta \ge {\bar \delta}$, the default probability   $P_d \le (1-\delta)$ and  $E[G] = 1-\delta$ and this is true for all $\epsilon$. Hence by part (a), $\epsilon_t \to 0$ almost surely.   \eop
 \end{enumerate}
 \end{cor}
\textbf{Remarks:}
{\it Thus from part (c), under the conditions of Corollary \ref{Lem_avg_dyn},
the random dynamics always converges to all `risky' agents (pure ESS), while 
the average dynamics, as given by Corollary \ref{Lem_avg_dyn},  either converges to pure or mixed ESS further based on   system parameters (mainly various  rates of return).} 
 
{\it From this partial analysis (corollaries are for  large enough $\delta$) it  appears  that one can never have mixed ESS with random dynamics, and this is  a big contrast to the average dynamics; when agents observe sparsely the network eventually settles to one of the two strategies, and  if they observe more samples there is a possibility of emergence of mixed ESS that balances the two returns.} We observe similar things, even for $\delta$ as small as $0.8$ in numerical simulations (Table \ref{MC _based avg and random}). We are keen to understand this aspect in more details as a part of the future work.

  To summarize we have a  financial network which grows with new additions, in which the new agents adapt one of the two available strategies based on the returns of the agents that they observed/interacted with.  Our asymptotic analysis of \cite{Systemicrisk} was instrumental in deriving these results. This is just an initial study of the topic.  One can think of other varieties of dynamics, some of which could be a part of our future work. The existing agents may change their strategies depending upon their returns and   observations. The agents might leave the network if they have reduced returns  repeatedly.  The network may adjust itself without new additions etc. 

\section{Numerical observations}
We performed  numerical simulations to validate our theory. We included Monte-Carlo (MC) simulation based dynamics in which the clearing vectors are also computed by directly solving the fixed point equations, for any given sample path of shocks. Our theoretical observation well matches the MC based limits. 

In Tables \ref{Average dynamics u_1  more than u_2},  \ref{Average dynamics u_1 close to  u_2}  and \ref{Average dynamics u_1 less than u_2} we tabulated the limits of the average dynamics for various scenarios, and the observations match 
the results of Corollary \ref{Lem_avg_dyn}. The configuration used for Table \ref{Average dynamics u_1  more than u_2}  is:   $n_0 = 2000, \epsilon_0 =0.75, r_s = 0.18, r_b= 0.19,  w= 100, v= 46,  \alpha =0.1$, while that used for Table \ref{Average dynamics u_1 close to  u_2}  is: $n_0 = 2000, \epsilon_0 =0.5, r_s = 0.17, r_b= 0.19,  w= 100, v= 40,  \alpha =0.1$. For both these tables risky expected rate of returns ${\bar r}_r$ is smaller than $r_b$ and the dynamics converges either to `all risky' agents configuration or to a mixed ESS. 
In Table \ref{Average dynamics u_1 less than u_2}, the risky expected rate of returns ${\bar r}_r = .1250 $ which is greater than $r_b$ and $r_s$, thus the dynamics converges to all risky-agents, as  indicated  by  Corollary \ref{Lem_avg_dyn}.

\begin{table}[]
 
\begin{minipage}{6cm}

\centering
\begin{tabular}{|l|l|l|l|l|l|}
\hline
 $u$& $d$ & $\delta$ & $\phi_1$  & $\phi_2$  & $\epsilon^{*}$  \\ \hline
 0.2& -0.05 &0.8  & 72 & 0 & 1 \\ \hline
 0.2& -0.1 &0.8  & 72 & 0 & 1 \\ \hline
 0.2& -0.15 &0.8  &72  & 0 & 1 \\ \hline
 0.2& -0.2 &0.8  &72  & 0 & 1 \\ \hline
 0.2& -0.25 &0.8  &72  & 0 & 1 \\ \hline
\end{tabular}
\vspace{4mm}
 \caption{When the shocks are too large along with larger taxes ($v = 46$),  the average dynamics converges to  a configuration   with all `risk-free agents'!}\label{Average dynamics u_1  more than u_2}
 
\end{minipage}
\hspace{0.3cm}
\begin{minipage}{6cm}
\vspace{-5mm}
\centering
\begin{tabular}{|l|l|l|l|l|l|}
\hline
 $u$& $d$ & $\delta$ & $\phi_1$  & $\phi_2$  & $\epsilon^{*}$  \\ \hline
 0.2& -0.1 &0.95  & 78.33 & 78.27 & 0.3326  \\ \hline
 0.2& -0.11 &0.95 & 78.24 & 78.31 & 0.3791 \\ \hline
 0.2& -0.12 &0.95  &78.14  & 78.14 & 0.4288 \\ \hline
 0.2& -0.13 &0.95  &78.04  & 78.04 & 0.4820\\ \hline
 0.2& -0.14 &0.95  &77.92  & 77.92 & 0.5385 \\ \hline
\end{tabular}
\vspace{4mm}
  \caption{ Average dynamics converges to mixed ESS, at which  both populations survive with  $\phi_1  =  \phi_2$, }\label{Average dynamics u_1 close to  u_2}
 \end{minipage}
 
 \begin{minipage}{4.5cm}
 \centering
\begin{tabular}{|l|l|l|l|l|l|}
\hline
 $u$& $d$ & $\delta$ & $\phi_1$  & $\phi_2$  & $\epsilon^{*}$  \\ \hline
 0.15& -0.1 &0.9  &0  & 82.12 & 0  \\ \hline
 0.16& -0.1 &0.9 & 0 & 83.24 & 0 \\ \hline
 0.17& -0.1 &0.9  &0  & 84.29 & 0 \\ \hline
 0.18& -0.1 &0.9  &0  & 85.19 & 0\\ \hline
\end{tabular}
\vspace{4mm} 
 \caption{ Average Dynamics converges to   all `risky-agents'; Configuration: $n_0 = 2000, \epsilon_0 =.5, r_s = 0.10, r_b= 0.12,  w= 100, v= 30,  \alpha =0.5$}\label{Average dynamics u_1 less than u_2}
 \end{minipage}
 \hspace{4mm}
\begin{minipage}{6.5cm}

\centering
\begin{tabular}{|c|c|c|c|c|c|}
\hline
\textbf{Config} & \multicolumn{2}{|c|}{\textbf{  $\epsilon^* $(Theory)}} & \multicolumn{2}{c|}{\textbf{$\epsilon^{\star}$(Monte Carlo)}} \\ \hline
\textbf{($d, \delta, v$)} &
\textbf{Avg}   & \textbf{Rndm} & \textbf{Avg} & \textbf{Rndm} \\ \hline
\ 0.10, \  0.95, \ 40 &0  & 0 &.0016 & 0.0011  \\ \hline

-0.10, \  0.95, \ 40 &0.33 & 0 &.3214  &0.0004  \\ \hline

-0.15, \ 0.95, \ 40  &0.6 & 0 & .5988  & 0.0014 \\ \hline

\ 0.10, \ 0.80, \ 46  & 1 & 0 & .9896  & 0.0065  \\ \hline
\end{tabular}
\vspace{4mm}
\caption{ Average   and Random dynamics,  Comparison  of MC results with theory   Configuration: $n_0 = 2	000, u = 0.2,   r_s = 0.17, r_b= 0.19,  w= 100,   \alpha =0.1$}\label{MC _based avg and random}
\end{minipage}
\vspace{-8mm}
\end{table}

In Table \ref{MC _based avg and random} we considered random dynamics as well as average dynamics. In addition, we provided the Monte-Carlo based estimates. There is a good match between the MC estimates and the theory. Further we have the following observations: a) random dynamics always converge to a configuration with all `risky' agents, as given by Corollary \ref{Cor_rand};    
 b) when ${\bar r}_r >  r_b$, the average dynamics also converges to $\epsilon^* =0$ as suggested by Corollary \ref{Lem_avg_dyn};
 and c)   when ${\bar r}_r <  r_b$, the average dynamics converges to mixed ESS or to a configuration with all `risk-free' agents, again as given by  Corollary \ref{Lem_avg_dyn}.

As the `risk increases',  i.e., as the amount of taxes increase and or as the expected rate of return of risky investments ${\bar r}_r$ decreases, one can observe that the average dynamics converges to all `risk-free' agents (last row of   Table  \ref{MC _based avg and random}) thus averting systemic risk event (when there are large number of defaults, $P_d$). While the random dynamics fails to do the same. 
As predicted by theory (the  configurations satisfying  part (b) of Corollary \ref{Cor_rand}), random dynamics might also succeed in averting the systemic risk event, when the expected number of defaults is one for all $\epsilon >0.$ It is trivial to verify that the configuration with $w(1+u) < v$, is one such example.  Thus, average dynamics is much more robust towards averting systemic risk events.

\section{Conclusions}

We consider a  financial network with a large number of agents. The  agents are interconnected via liability graphs.  
There are two types of agents, one group lends to others and invests the rest in risk-free projects, while the second group borrows/lends and invests the rest in risky ventures. 
Our study is focused on analysing the emergence of these groups, when the new  agents adapt their strategies for the next investment round based on the returns of the previous round. 
We considered two types of dynamics; in average dynamics the new agents observe large sample of data before  deciding their strategy, and in random dynamics  the decision is based on a small random sample. 

We have the following important observations: a) when the expected rate of return of the risky investments is higher (either when the shocks are rare or when the shocks are not too large) 
than the risk-free rate, then `risk-free' group wipes out eventually, almost all agents go for risky ventures;   this is true  for both types of dynamics;  
b) when the expected rate  of  risky investments is smaller,  a mixed ESS can emerge  with average dynamics while  the random dynamics always converges to all risky agents;  at mixed ESS the expected returns of both the groups are equal;  more interestingly, when the risky-expected rate is too small, the average dynamics converges to a configuration with  all risk-free agents. 

 In other words,
in scenarios with  
  possibility of a systemic risk event, i.e., when there is a possibility of the complete-system collapse (all agents default),  the average dynamics manages to wipe out completely the risky agents; the random dynamics can fail to do the same.   Thus when agents make their choices rationally and after observing sufficient sample of the returns of the previous round of investments, there is a possibility to avoid systemic risk events.  These are some  initial results and we would like to investigate further  in future to make more affirmative statements in this direction.

 \bibliographystyle{plain}

\begin{thebibliography}{10}


\bibitem{Systemicrisk} 
 Veeraruna Kavitha, Indrajit Saha, and Sandeep Juneja. 
\newblock  "Random Fixed Points, Limits and Systemic risk." 
\newblock {\em In 2018 IEEE Conference on Decision and Control (CDC)}, pp. 5813-5819. IEEE, 2018.
\bibitem{ESS}Miekisz, Jacek. "Evolutionary game theory and population dynamics." Multiscale Problems in the Life Sciences. Springer, Berlin, Heidelberg, 2008. 269-316.

\bibitem{Benven} Benveniste, Albert, Michel Métivier, and Pierre Priouret. Adaptive algorithms and stochastic approximations. Vol. 22. Springer Science \& Business Media, 2012.

\bibitem{eisenberg2001systemic}
Larry Eisenberg and Thomas~H Noe.
\newblock Systemic risk in financial systems.
\newblock {\em Management Science}, 
2001.
\bibitem{acemoglu2015systemic}
Daron Acemoglu, Asuman Ozdaglar, and Alireza Tahbaz-Salehi.
\newblock Systemic risk and stability in financial networks.
\newblock {\em The american economic review}, 
2015.

\bibitem{allen2000financial}
Franklin Allen and Douglas Gale.
\newblock Financial contagion.
\newblock {\em Journal of political economy},  
2000.
\bibitem{LiHonggang} Li, Honggang, Chensheng Wu, and Mingyu Yuan. "An evolutionary game model of financial markets with heterogeneous players." Procedia Computer Science 17 (2013): 958-964.
\bibitem {YangKe}Yang, Ke, Kun Yue, Hong Wu, Jin Li, and Weiyi Liu. "Evolutionary Analysis and Computing of the Financial Safety Net." In International Workshop on Multi-disciplinary Trends in Artificial Intelligence, pp. 255-267. Springer, Cham, 2016.

\bibitem{Brock}Brock, William A., and Cars H. Hommes. "Heterogeneous beliefs and routes to chaos in a simple asset pricing model." Journal of Economic dynamics and Control 22, no. 8-9 (1998): 1235-1274.
\bibitem{Friedman} Friedman, Daniel. "Towards evolutionary game models of financial markets." (2001): 177-185.
\bibitem{Borkar}Borkar, Vivek S. Stochastic approximation: a dynamical systems viewpoint. Vol. 48. Springer, 2009.

\bibitem{MeanWireless} Tembine, Hamidou, Eitan Altman, Rachid El-Azouzi, and Yezekael Hayel. "Evolutionary games in wireless networks." IEEE Transactions on Systems, Man, and Cybernetics, Part B (Cybernetics) 40, no. 3 (2009): 634-646.
\bibitem{Smith} Smith, J. Maynard, and George R. Price. "The logic of animal conflict." Nature 246, no. 5427 (1973): 15-18.
 
\end{thebibliography}

\TR{
\vspace{-7.5mm}
\section*{Appendix }
\noi \textbf{Proof of Lemma \ref{Lem_Average_clearing}:}  We consider the following all possible cases: \\
\underline{Case 1:} First consider the case when downward shock can be absorbed, in this case the clearing vector  ${\bar x}^{ \infty} = y \delta + y(1-\delta) =y$, default probability is  $ P_d = 0$. The region is true if the following condition is meet i.e.,  if
\begin{equation}
  k_d-v + yc_{\epsilon} > y \implies c_{\epsilon} > \frac{y-\underline{w}}{y}.\nonumber
  \end{equation}
\noindent \underline{Case 2:} Consider the case with banks receive shock will default and the 
 corresponding average clearing vector  ${\bar x}^{ \infty} = y\delta+ (\underline{w} +c_\epsilon {\bar x}^{ \infty})(1-\delta) $ which simplifies to:
 \begin{equation}
{\bar x}^{ \infty} = \frac{y\delta+ \underline{w}(1-\delta) }{1-c_\epsilon(1-\delta)}. \nonumber
\end{equation}
This region lasts if the following conditions hold  to be true 

$$ k_d- v +c_{\epsilon} {\bar x}^{ \infty} <y, \mbox{ and  } k_u- v +c_{\epsilon} {\bar x}^{2 \infty} > y.$$
\noindent Substituting  ${\bar x}^{ \infty} = \frac{y\delta+ \underline{w}(1-\delta) }{1-c_\epsilon(1-\delta) }$  we have, \vspace{-12mm}
\begin{equation}
\hspace{56mm}
\frac{y-\overline{w}}{y-(1-\delta)(\overline{w}-\underline{w})} <
c_\epsilon <\frac{y-\underline{w}}{y}\nonumber.
\end{equation}

\noindent \underline{Case 3:}
 In this we first calculate ${\bar x}^{ \infty}$ which is obtained by solving following fixed point equation:
  \begin{eqnarray*}
   {\bar x}^{ \infty} &= & (k_d- v+ c_\epsilon {\bar x}^{ \infty})(1-\delta ) + (k_u- v+ c_\epsilon {\bar x}^{ \infty})\delta  \ \ 
    \implies {\bar x}^{ \infty} \ =  \  \frac{E W}{1-c_\epsilon}.
   \end{eqnarray*}
In this case the default probability is $P_d = 1$.
The regime satisfies if the following hold
\begin{eqnarray*}
   k_u -v + c_\epsilon \frac{E W}{1-c_\epsilon} < y \ \ 
   \implies c_{\epsilon }< \frac{y- \overline{w}}{y- (1-\delta)(\overline{w}-\underline{w})}. \hspace{5mm}  \mbox{ \eop }
    \end{eqnarray*}
      
\noindent{\bf Proof of Corollary \ref{Lem_avg_dyn}:}  First consider the system with $\delta = 1$, i.e., system without shocks. From Lemma \ref{Lem_Average_clearing}, $P_d  \le (1-\delta)$ for all 
$\epsilon$  because (with $\delta = 1$)
\vspace{-2mm}

{\small
 $$
y \left (\hspace
{-1mm}c_\epsilon - \frac{ y - {\bar w} }{y } \right )
=   w (1+\epsilon) (1+u) - v -  w \epsilon(1+r_b) = w (1+u) - v  + w \epsilon (u - r_b) \ge w (1+u) - v,
$$} 
for all $\epsilon$ (the lower bound independent of $\epsilon$). Under these assumptions,  there exists ${\bar \delta} < 1$ by continuity of the involved functions such that 
$$
y \left ( c_\epsilon - \frac{ y - {\bar w} }{y - (1-\delta) ({\bar w} - {\underline  w})  } \right )  > 0  \mbox{ for all }  \delta \ge {\bar \delta} \mbox{ and for all } \epsilon.
$$
Thus from Lemma \ref{Lem_Average_clearing} ${\bar x}^\infty = y$ or ${\bar x}^\infty =  \frac{\delta y +(1-\delta) \underline{w}}{1-(1-\delta)c_\epsilon}$ for all such $\delta \ge {\bar \delta}$.  We would repeat a similar trick again, 
so assume initially  ${\bar x}^\infty = y$ for all $\epsilon$ and consider  $\delta \ge {\bar \delta}$.  With this assumption we  will  have:
\begin{eqnarray}
\label{eqn_R1eps}
R^1 (\epsilon)  &=&  \left (  w\epsilon(1+r_s)+ \frac{(1-\alpha)(1-\epsilon)}{(\alpha +\epsilon)}  y  -v  \right )^+  \\
&= &   \left (  w\epsilon(1+r_s)+  w (1-\epsilon) (1+r_b)  -v  \right )^+ \nonumber  \\
&=& \left (    w (1+r_b) - v  + w \epsilon (r_s - r_b)   \right )  \nonumber
 \mbox{, under the given hypothesis,  and }  \\
R^2 (\epsilon) &=&    \left (K_i+\frac{\alpha (1+\epsilon) }{\alpha + \epsilon}   y -v-y \right )^+ 
=    \left ( K_i  -  w \epsilon (1+r_b)  -v  \right )^+  \label{eqn_R2eps}  \\
& = & \left \{  
\begin{array}{llll}
R^2_u 
  &  \mbox{ w.p. }  \delta  &  \mbox{ where  }  R^2_u  := w (1+u) - v + w \epsilon (u-r_b)   \\
\left (  R^2_d   \right )^+ &  \mbox{ w.p. } 1-  \delta  &   \mbox{ where  }  R^2_d  :=    w (1+d) - v + w \epsilon (d-r_b)   . 
\end{array} \right .  \nonumber
\end{eqnarray}
Note that $R^2_u \ge w (1+u) - v > 0$ (for any $\epsilon$) under the given hypothesis. \\
{\bf Proof of part (a):}  When ${\bar r}_r  > r_b$, from \eqref{Eqn_phis}, it is clear that (inequality only when $R^2_d $ is negative)
\begin{eqnarray*}
\phi_2 (\epsilon) - \phi_1 (\epsilon)   \ge  R^2_u \delta + R^2_d (1-\delta) - \phi_1 (\epsilon)   = w ({\bar r}_r - r_b) + w \epsilon ({\bar r}_r - r_s)   > 0.
\end{eqnarray*}  
Thus in this case $\phi_2 > \phi_1$ for all $\epsilon$ and hence 
$$
g(\epsilon) < 1/2  \mbox{ and }  2 g(\epsilon) - 1 < 0 \mbox{ for all $0< \epsilon < 1$.  }
$$
Therefore with Lyaponuv function $V_0(\epsilon) = \epsilon  / (1-\epsilon) $ on defined on  neighbourhood $[0, 1)$ of $0$ (in relative   topology on $[0, 1]$) we observe that
$$
\frac{dV_0}{d \epsilon}  h (\epsilon)  =   \frac {\epsilon}{1-\epsilon}     ( 2 g(\epsilon) - 1 )  <  0  \mbox{ for all } 0 <  \epsilon  < 1  \mbox{ and equals } 0 \mbox{ for } \epsilon = 0.
$$ 
Further $V_0(\epsilon)  \to \infty$ as $\epsilon \to 1$, the boundary point of $[0, 1)$. 
Thus $\epsilon^* = 0$ is the asymptotically stable attractor of ODE (\ref{Eqn_Avg_ODE})  (see \cite[Appendix, pp.148]{Borkar}) and hence the result  follows by Theorem \ref{Thm_avg_conv}.

For all $\delta \ge {\bar \delta}$, from Lemma \ref{Lem_Average_clearing}, we have   the following 
\begin{eqnarray}
\sup_{\epsilon}|y-{\bar x}^\infty| = \sup_{\epsilon} (1-\delta) \bigg |\frac{y-yc_\epsilon -\underline{w}}{1-(1-\delta)c_\epsilon}\bigg| <  \frac{1-\delta}{\delta} \eta
\end{eqnarray}
for some $\eta > 0$ , which decreases to 0 as  $\delta \to 1$. (The last inequality is due to $c_\epsilon < 1$ and then taking  supremum over  $\epsilon$). 
By continuity of  the above upper bound with respect to $\delta$  
and the subsequent functions considered in the above parts of the proof, there exists a ${\bar \delta} < 1$ (further big if required) such that all the above arguments are true for all $\delta > {\bar \delta}$. 

\noi{\bf Proof of part (b):}  The proof follows in similar way, now using Lyaponuv function $V_1(\epsilon) = (1- \epsilon) / \epsilon$ on neighbourhood $(0, 1]$ of 1,  and  by observing that $g(\epsilon )  > 1/2$ for all $\epsilon < 1$ and hence
$$
\frac{dV_1}{d \epsilon}  h (\epsilon)  =  -   \frac {1-\epsilon}{\epsilon}     ( 2 g(\epsilon) - 1 )  <  0  \mbox{ for all } 0 <  \epsilon  < 1  \mbox{ and equals } 0 \mbox{ for } \epsilon = 1.
$$ 
 
\noi{\bf Proof of part (c):}  It  is clear that  $\phi_1 (\epsilon) = R^1(\epsilon) $   decreases linearly as $\epsilon$ increases:  
$$
\phi_1 (\epsilon) =  w(1+ r_b) - v +  w\epsilon  (r_s  - r_b).
$$  

For   $\epsilon$ in the neighbourhood of $0$, $\phi_2 (\epsilon) >0$ and is decreasing linearly with slope ${\bar r}_r - r_b$, because $R^2_d (0) =    w(1+ d) - v  > 0$ and thus for such $\epsilon$
$$
\phi_2 (\epsilon) =  w(1+{\bar r}_r) - v +  w\epsilon  ({\bar r}_r - r_b).
$$
From \eqref{eqn_R2eps}, $R^2_d ({ \epsilon})$ is decreasing with increase in $\epsilon.$
  There is a  possibility   of an ${\bar \epsilon}$ 
that satisfies  $R^2_d ({\bar \epsilon})  =  0$, in which case $\phi_2$  increases  linearly with slope $\delta w (u - r_b)$, i.e., 
$$
\phi_2 (\epsilon) = \delta \left [   w(1+ u)  - v +  w\epsilon  ( u - r_b) \right ]  \mbox{ for all } \epsilon \ge {\bar \epsilon}.
$$
 
When ${\bar r}_r < r_b$ we have, 
$$\hspace{24mm}\phi_1 (0) = w(1+r_b) - v > w(1+{\bar r}_r) - v = \phi_2 (0).$$ 
By hypothesis $\phi_1 (\epsilon) < \phi_2 (\epsilon)$ for some $\epsilon$, hence  by intermediate  value theorem there exists at least one $\epsilon^*$ that satisfies $\phi_1 (\epsilon^*) = \phi_2 (\epsilon^*).$
Further the zero is unique because $\phi_2$ is either linear or piece-wise linear (with different slops), while $\phi_1 $ is linear.

Consider Lyaponuv function  $V_*(\epsilon) := (\epsilon - \epsilon^*)^2 /( \epsilon (1-\epsilon) )$ on neighbourhood $(0,1)$ of $\epsilon^*$, note $V_*(\epsilon) \to \infty$ as $\epsilon \to 0$ or $\epsilon \to 1$  and observe by (piecewise) linearity of the functions we will have 
\begin{eqnarray*}
\phi_1 (\epsilon) & >& \phi_2 (\epsilon)  \mbox{ and thus }  (2 g(\epsilon) - 1) >  0  \mbox{ for all } 0 <  \epsilon < \epsilon^*  \mbox{ and }
\\  
\phi_2(\epsilon) & >&  \phi_1 (\epsilon)   \mbox{ and thus }  (2 g(\epsilon) - 1) < 0   \mbox{  for all }  1 > \epsilon > \epsilon^*. 
    \end{eqnarray*} Thus we have\footnote{When $\epsilon < 1/2$  and $\epsilon < \epsilon^*$ then clearly $ \frac {  (\epsilon - \epsilon^*)   (2\epsilon - 1)  } { \epsilon (1-\epsilon) }    > 0$. 
    When $\epsilon > 1/2$ we have $(2 \epsilon - 1)  / \epsilon <  1/2$  and with   $\epsilon < \epsilon^*$ we have $\epsilon^* - \epsilon    < 1 - \epsilon$  and  thus 
    $$
    2  +  \frac {  (\epsilon - \epsilon^*)  (2\epsilon - 1)  } { \epsilon (1-\epsilon) }  \ge  3/ 2 > 0 \mbox { for all } \epsilon < \epsilon^*. 
    $$
    In a similar way $\epsilon > \epsilon^*$, then we will have that the above term is again positive.
    },
$$
\frac{dV_*}{d \epsilon}  = 2  \frac{\epsilon -\epsilon^* } {\epsilon (1-\epsilon) } +    \frac{ (\epsilon -\epsilon^*)^2 (2\epsilon - 1) }   {\epsilon^2 (1-\epsilon)^2 } \mbox{ and hence }
$$

$$
\frac{dV_*}{d \epsilon}  h (\epsilon)  =   (\epsilon - \epsilon^*)    \left ( 2  +  \frac {  (\epsilon - \epsilon^*)  (2\epsilon - 1)  } { \epsilon (1-\epsilon) }  \right ) ( 2 g(\epsilon) - 1 )  <  0  \mbox{ for all }  \epsilon \notin  \{0, 1, \epsilon^*\}.
$$ 
Thus $\epsilon^* $ is the asymptotically stable attractor of ODE (\ref{Eqn_Avg_ODE}) and hence the result follows  by Theorem \ref{Thm_avg_conv}.
The result can be extended for $\delta < 1$ as in case (a) and 
the rest of the details follow by direct verification (at $\delta = 1$), i.e., by showing that $\phi_1(\epsilon^*) = \phi_2(\epsilon^*)$ at $\delta = 1$ and the equality is satisfied approximately in the neighbourhood of $\delta =1$.  \eop

\noindent {\bf Proof of Corollary \ref{Lem_random_dyn}:}  For part (a), 
  $h_R (\epsilon) = - c_G \epsilon (1-\epsilon)$, where the constant $c_G = 1$ (or respectively $c_G = 2 \delta - 1$). Using Lyanponuv function of part (a) of Corollary \ref{Lem_avg_dyn}, the proof follows in exactly the same lines.
  
\noindent  For part (b),  $h_R (\epsilon) =  \epsilon (1-\epsilon)$, and proof follows as in part (b) of Corollary \ref{Lem_avg_dyn}.
For part  (c), first  observe 
(using equations \eqref{eqn_R1eps}-\eqref{eqn_R2eps} of proof of  Corollary \ref{Lem_avg_dyn})
\begin{eqnarray*}
 R^2_u  (\epsilon) -  R^1 (\epsilon) 
& \geq & w(1+u) +w\epsilon(u- r_s)-y +  {\bar x}^\infty\bigg(\frac{2\alpha +\epsilon -1}{\alpha +\epsilon}\bigg) \\
 &= & w(1+u) +w\epsilon(u- r_s) +  ({\bar x}^\infty -y) -{\bar x}^\infty \bigg( \frac{1-\alpha}{\alpha +\epsilon}\bigg) \\
  &=&  w(u-r_b) +w\epsilon(u- r_s) +  ({\bar x}^\infty -y)\bigg(1- \frac{1-\alpha}{\alpha +\epsilon}\bigg) >  0. 
 \end{eqnarray*}
 The last inequality is trivially true for $\delta = 1$ (and so ${\bar x}^\infty=y$) for the given hypothesis, and then by continuity as in proof of Corollary \ref{Lem_avg_dyn},  one  can consider $\bar{\delta} <1$   such that  for all  $\delta    \geq \bar{\delta}$, the term $({\bar x}^\infty -y)  \big(1- \frac{1-\alpha}{\alpha +\epsilon}\big)$ (uniformly over $\epsilon$) can be made arbitrarily small.  
 When $P_d = 0$, i.e., ${\bar x}^\infty=y$ for some $\epsilon$, then 
 $
 R^2_d (\epsilon) - R^1 (\epsilon) = w(d-r_b) + w\epsilon (d-r_s) < 0$ for all such $\epsilon$. 
 When $P_d \ne 0$, then $R^2_d = 0  \le R^1$. Thus in either case 
 $
 R^2_d (\epsilon) \le R^1 (\epsilon)$ for all $\epsilon$.    
 
By  virtue of the above  arguments we have  $P_d \le (1-\delta)$ and  $E[G] = 1-\delta$ and this is true for all $\epsilon$, for all $\delta \ge {\bar \delta}$. The rest of the proof follows from part(a).
  
\eop 
 
\end{document}